\documentclass[preprint,showpacs,showkeys,preprintnumbers,amsmath,amssymb,prb]{revtex4}

\usepackage{graphicx}
\usepackage{float}

\begin{document}

\preprint{Winterlik et al., Rb$_4$O$_6$ and Cs$_4$O$_6$, PRB}

\title{Exotic magnetism in the alkali sesquoxides Rb$_4$O$_6$ and Cs$_4$O$_6$}

\author{J{\"u}rgen Winterlik, Gerhard~H. Fecher, Taras Palasyuk, Claudia Felser}
\affiliation{Institute for Inorganic and Analytic Chemistry, Johannes Gutenberg - Universit\"at, D-55099 Mainz, Germany.}

\author{J{\"u}rgen K{\"u}bler}
\affiliation{Institute for Solid State Physics, Technische Universit\"at, 64289 Darmstadt, Germany}

\author{Claus M{\"u}hle, Martin Jansen}
\affiliation{Max - Planck - Institute for Solid State Research, 70569 Stuttgart, Germany} 

\author{Ivan Trojan, Sergey Medvedev, Mikhail~I. Eremets}
\affiliation{Max - Planck - Institute for Chemistry, 55020 Mainz, Germany}

\author{Franziska Emmerling}
\affiliation{Bundesanstalt Materialforschung und Pr{\"u}fung Berlin, 12489 Berlin, Germany}

\date{\today}

\begin{abstract}

Among the various alkali oxides the sesquioxides Rb$_4$O$_6$ and Cs$_4$O$_6$
are of special interest. Electronic structure calculations using the local 
spin-density approximation predicted that Rb$_4$O$_6$ should be a 
half-metallic ferromagnet, which was later contradicted when an experimental 
investigation of the temperature dependent magnetization of Rb$_4$O$_6$ showed 
a low-temperature magnetic transition and differences between
zero-field-cooled (ZFC) and field-cooled (FC) measurements. Such behavior is
known from spin glasses and frustrated systems. Rb$_4$O$_6$ and Cs$_4$O$_6$ 
comprise two different types of dioxygen anions, the hyperoxide and the peroxide 
anions. The nonmagnetic peroxide anions do not contain unpaired electrons while
the hyperoxide anions contain unpaired electrons in antibonding $\pi^*$-orbitals. 
High electron localization (narrow bands) suggests that electronic correlations are of major
importance in these open shell $p$-electron systems. Correlations and charge ordering 
due to the mixed valency render $p$-electron-based anionogenic  
magnetic order possible in the sesquioxides. In this work we present an 
experimental comparison of Rb$_4$O$_6$ and the related Cs$_4$O$_6$. 
The crystal structures are verified using powder x-ray diffraction.
The mixed valency of both compounds is confirmed using Raman spectroscopy, and
time-dependent magnetization experiments indicate that both compounds show 
magnetic frustration, a feature only previously known from $d$- and $f$-electron systems.

\end{abstract}



\maketitle

\section{Introduction}

Magnetism arising from $d$- and $f$-shells has drawn the bulk of research attention but 
$p$-electron based magnetic order is a rare and fascinating topic that presents the added 
challenge of molecular and not just atomic ordering. The majority of main group molecules are nonmagnetic.
Few exceptions are found, e.g. in NO, NO$_2$ and O$_2$. Molecular oxygen contains two single electrons 
in degenerate antibonding $\pi^*$-orbitals, which can order magnetically in a solid crystal.
Solid oxygen shows a large variety of magnetic phenomena ranging from antiferromagnetism to 
superconductivity.\cite{FrJ04,MeH84,GMU04,AKH95,SSI98}

What applies to molecular oxygen applies equally to charged oxygen molecules with unpaired 
electrons. Dioxygen anions are principally found in alkali and alkaline earth oxides, which 
represent excellent model systems because of their supposedly simple electron configurations. 
The hyperoxide anion O$_2^-$ corresponds to "charged oxygen". Since it still contains one 
unpaired electron, magetic order is enabled for hyperoxides. KO$_2$, RbO$_2$, and 
CsO$_2$ are known to exhibit antiferromagnetic ordering below their respective N{\'e}el 
temperatures of 7~K, 15~K, and 9.6~K, respectively.\cite{LRK79,HJS89}

Among the alkali oxides, the sesquioxides are of special interest. In contrast to related 
compounds, which are white, yellow, or orange, the sesquioxides Rb$_4$O$_6$ and Cs$_4$O$_6$ 
are black. Furthermore, a formula unit AM$_4$O$_6$ (AM~=~alkali metals Rb or Cs) contains two 
different types of dioxygen anions: one closed-shell nonmagnetic peroxide anion and two of 
the aforementioned hyperoxide anions. The structural formula of the sesquioxides is thus accurately
represented as (AM$^+$)$_4$(O$_2^-$)$_2$(O$_2^{2-}$).\cite{JaK91} The mixed valency enables
complicated magnetic structures in the sesquioxides. The first descriptions of the crystal structures 
of the sesquioxides were published in 1939 in the pioneering works of Helms and 
Klemm.\cite{HeK391,HeK392,HeK393} Both compounds belong to the Pu$_2$C$_3$ structure type and 
to space group $I\:\overline{4}3d$. This Pu$_2$C$_3$ structure type is known from the 
noncentrosymmetric rare earth metal sesquicarbide superconductors such as Y$_2$C$_3$ with a maximum 
critical temperature of $T_c=18$~K.\cite{KSA08} For Cs$_4$O$_6$, the literature runs out 
after 1939 because of the extremely challenging synthesis and sensitivity to air. 
In Rb$_4$O$_6$, the presence of both peroxide and hyperoxide anions was verified
by neutron scattering\cite{JHK99}, and electronic structure calculations
using the local spin density approximation (LSDA) were performed to explain the exceptional 
black color.\cite{ADB05} These same calculations predicted a half-metalllic 
ferromagnetic ground state but were contradicted by later experiments in which a transition 
was found to occur at approximately 3.4~K in the temperature dependent 
magnetization.\cite{WFF07} Differences between ZFC and FC measurements indicated
that Rb$_4$O$_6$ behaves like a frustrated system or a spin glass. Recently
published electronic structure calculations are consistent with the 
 experimental findings.\cite{WFJ09} It was shown that
for an accurate theoretical treatment of highly localized systems such as Rb$_4$O$_6$
and Cs$_4$O$_6$, a symmetry reduction, exact exchange and electron-electron correlations 
have to be considered in the calculations. This can be generalized to
any other open shell system that is based on $p$-electrons. Accounting for these 
features the calculations result in an insulating ground state. 
Further calculations using the spin spiral method show that Rb$_4$O$_6$
exhibits spin spiral behavior in a certain crystal direction. The energy changes are extremely 
small along this direction indicating a multidegenerate ground state.\cite{WFJ09} 
In this work we present the routes of synthesis for Rb$_4$O$_6$ and Cs$_4$O$_6$ 
and the structural verification using powder x-ray diffraction (XRD).
Raman spectroscopic measurements confirm the mixed valency of the used 
Rb$_4$O$_6$ and Cs$_4$O$_6$ samples. Magnetization experiments are shown
that indicate a dynamic time dependent magnetism of Rb$_4$O$_6$. Furthermore 
we present a comprehensive experimental study of Cs$_4$O$_6$, which has not been
extensively investigated due to the difficulty of sample preparation. 
Magnetization measurements dependent on temperature, magnetic field, and
time provide evidence that Cs$_4$O$_6$ shows a similar behavior as Rb$_4$O$_6$.
According to experimental and theoretical investigations,\cite{WFJ09} both 
sesquioxides are mixed valent highly correlated systems exhibiting
$p$-electron based magnetic frustration. These seemingly simple compounds can serve as model systems
for any other open-shell systems that are based on $p$-electrons such as hole-doped
MgO~\cite{ERC07} or nanographene.\cite{SCL06}

\section{Structure} 

\begin{figure}[H]
\includegraphics[width=9cm]{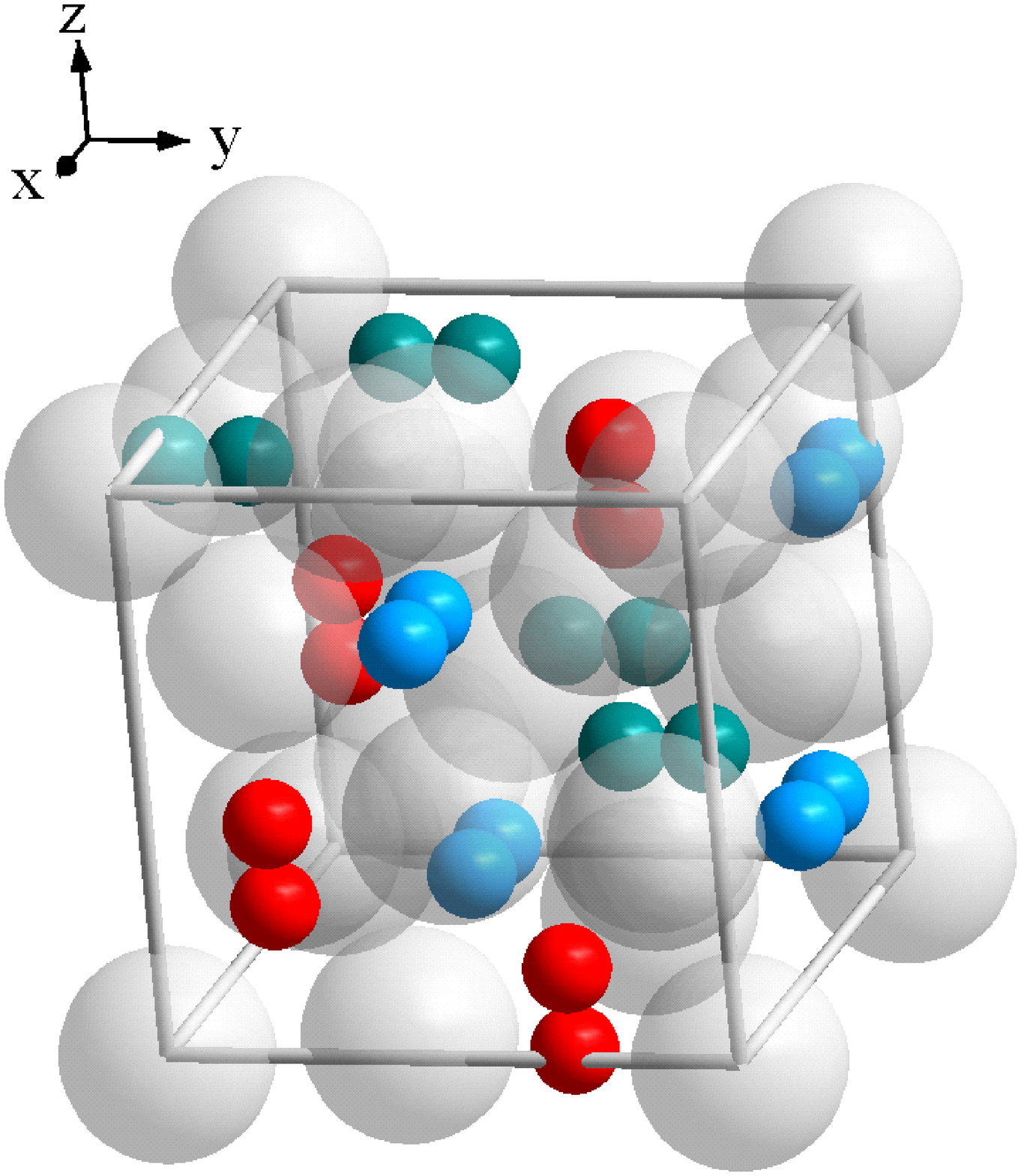}
\caption{(Color online) Pseudo-body-centered cubic cell of the alkali sesquioxides
AM$_4$O$_6$ (AM = Rb, Cs). Differently oriented dioxygen anions are drawn with
different colors so that the alignment along the axes can be
distinguished. For clarity, the Rb/Cs atoms are gray and
transparent.} 
\label{fig_str}
\end{figure}

Figure~\ref{fig_str} depicts a body-centered cubic unit cell of AM$_4$O$_6$. 
The 24 oxygen atoms in the cubic cell form 12 molecules that can be 
distinguished both by their valency and by their alignment along the principal 
axes. We assumed the nonmagentic peroxide anions to be oriented
along the $z$-axis, whereas the hyperoxides are oriented along the $x$- and
$y$-axes. The experimental bond lengths for the hyperoxide and the peroxide anions are
0.144~$a$ and 0.165~$a$, respectively\cite{BrJ92,SeJ98}. The cubic
lattice parameters are found in Section~\ref{sec_sc}.

\section{Synthesis}

The precursors rubidium and cesium oxide AM$_2$O as well as rubidium and cesium 
hyperoxide AMO$_2$ were prepared from elemental sources. For AM$_2$O, liquid rubidium/cesium,
purified by distillation, was reacted with a stochiometric amount of dry oxygen
in an evacuated glass tube followed by heating at 473~K for two weeks
under argon atmosphere.\cite{Hac28,Bra78} The samples were subsequently ground under argon
and the entire cycle was repeated five times. A slight excess of rubidium/cesium was distilled at
573~K in vacuum, resulting in a pale green powder of Rb$_2$O and an orange powder of Cs$_2$O. AMO$_2$ 
were prepared through the reaction of liquid rubidium/cesium and an excess of dry oxygen 
using the same method as described for AM$_2$O, resulting in yellow powders of RbO$_2$ and 
CsO$_2$, respectively. 

Rb$_4$O$_6$ was obtained by a solid state reaction of 400~mg RbO$_2$ 
and 160~mg Rb$_2$O in a molecular ratio of 4~:~1 in a glass tube sealed under argon at 
453~K for 24~hours. Cs$_4$O$_6$ was obtained in the same way using the amounts of 468~mg CsO$_2$ and 280~mg Cs$_2$O
and annealing at 473~K for 24~hours. For the reaction, a slight excess of Cs$_2$O 
was used to compensate for small amounts of cesium peroxide, which is always contained as an imurity
in Cs$_2$O. The very air sensitive products were ground and the reaction was repeated until 
pure phases were obtained (black powders). For the magnetization measurements, Rb$_4$O$_6$ 
and Cs$_4$O$_6$ were sealed in a high-purity quartz tube (Suprasil® glass) under helium atmosphere.

\section{Structural characterization}
\label{sec_sc}

\begin{figure}[H]
\centering
\includegraphics[width=9cm]{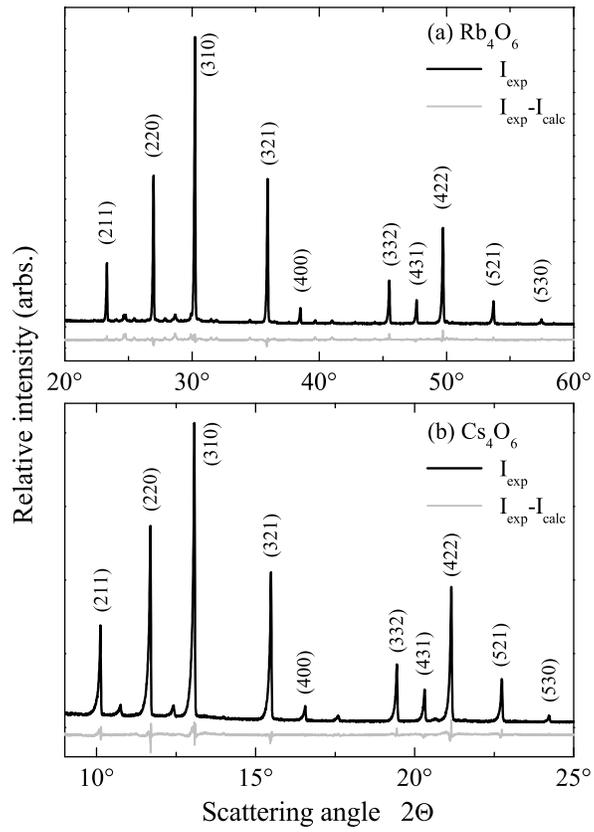}
\caption{Powder x-ray diffraction patterns of Rb$_4$O$_6$ and Cs$_4$O$_6$
         at 300~K (black). The gray curves show the differences between the
         observed data and the Rietveld refinements.}
\label{fig_xrd}
\end{figure}

The crystal structures of the compounds were investigated using XRD. The measurements 
were carried out using a Bruker D8 diffractometer with Cu K$_{\alpha1}$ radiation for Rb$_4$O$_6$
and Mo K$_{\alpha1}$ for Cs$_4$O$_6$. The samples were measured in sealed capillary tubes 
under an argon atmosphere. The diffraction patterns are shown in Figure~\ref{fig_xrd}.
The raw data (black) are compared to the difference between a calculated Rietveld refinement
and the raw data (gray). The refinements yielded weighted profile R-values of $R_{wp}=8.216$ 
for Rb$_4$O$_6$ and 6.388 for Cs$_4$O$_6$. Both compounds crystallize in the cubic structure $I\:\overline{4}3d$ 
(space group 220). The experimental lattice parameters as found from Rietveld refinements
are 9.322649(74)~\AA~for Rb$_4$O$_6$ and 9.84583(11)~\AA~for Cs$_4$O$_6$. The atomic parameters 
for both compounds are shown in Table~\ref{tab_xrd}, additional structural information is
found in Table~\ref{tab_qpara}. The pattern of Rb$_4$O$_6$ indicates good phase purity. 
In the case of Cs$_4$O$_6$, additional signals were detected. The signals were identified 
to arise from the impurity CsO$_2$, which belongs to the cubic space group 
$Fm\:\overline{3}m$ and has a lattice parameter of $a=6.55296(37)$~\AA. The impurity phase was included in
the refinement of Cs$_4$O$_6$. An impurity content of approximately 7.23\% of CsO$_2$ in Cs$_4$O$_6$
was derived from the refinement.

\section{Raman Spectroscopy}
\label{sec_rs}

\begin{figure}[H]
\includegraphics[width=9cm]{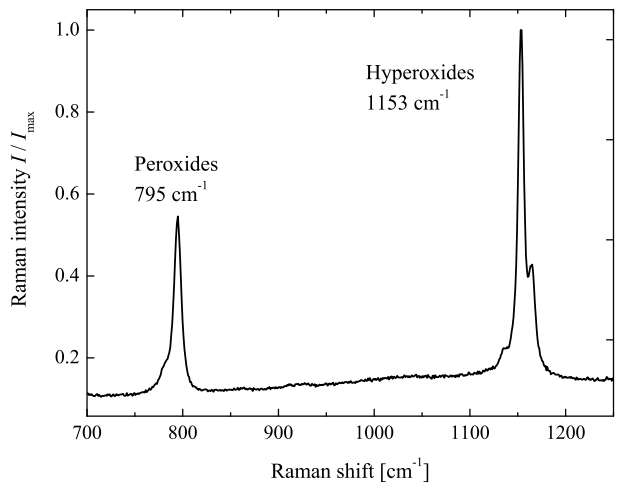}
\caption{(Color online) Raman spectrum of Rb$_4$O$_6$. The peak at 795~cm$^{-1}$
corresponds to the stretching vibration of the peroxide anions and the peak
at 1153~cm$^{-1}$ to the stretching vibration of the hyperoxide anions.} 
\label{fig_ramRb}
\end{figure}

The presence of peroxide and hyperoxide anions in Rb$_4$O$_6$ and Cs$_4$O$_6$ was verfied 
using Raman spectroscopy. The measurements were performed using a diamond anvil cell (DAC) 
in order to prevent sample decomposition from contact with air and moisture. 
The samples were loaded in a dry box under dry nitrogen atmosphere. The samples were confined 
within a cylindrical hole of 100 microns diameter and 50 microns height drilled in a Re 
gasket. We used synthetic type-IIa diamond anvils, which have 
only traces of impurities (\textless 1~ppm) and very low intrinsic luminescence. 
Raman spectra were recorded with a single 460-mm-focal-length imaging spectrometer (Jobin Yvon HR 460) 
equipped with 900 and 150 grooves/mm gratings, giving a resolution of 1–5~cm$^{-1}$, notch-filter 
(Kaiser Optics), liquid nitrogen cooled charge-coupled device (CCD (Roper Scientific)). 
Scattering calibration was done using Ne lines with an uncertainty of $\pm1$~cm$^{-1}$. The He-Ne laser 
of Melles Griot with the wavelength of 632.817 nm was used for excitation of the sample. The probing 
area was a spot with 5 microns diameter. Figure~\ref{fig_ramRb} shows the Raman spectrum
of Rb$_4$O$_6$ in a range from 700-1250~cm$^{-1}$. Two peaks are found at Raman shifts of 
795~cm$^{-1}$ and 1153~cm$^{-1}$, respectively. The signal at 795~cm$^{-1}$ corresponds to the 
stretching vibration of the peroxide anions and is in good agreement with the literature value 
of 782~cm$^{-1}$ for Rb$_2$O$_2$.\cite{EyT75} The peak at 1153~cm$^{-1}$ is assigned to the 
corresponding vibration of the hyperoxide anions and comparable to the literature value of 
1140~cm$^{-1}$ for RbO$_2$.\cite{BBB72} In the Raman spectrum for Cs$_4$O$_6$, which was
recorded as described above, signals of 738~cm$^{-1}$ and 1128~cm$^{-1}$ were recorded. 
The simultaneous presence of both dioxygen anion types and thus
the mixed valency is proven for both Rb$_4$O$_6$ and Cs$_4$O$_6$. 

\section{Magnetization}
\label{sec_mag}

The magnetic properties of Rb$_4$O$_6$ and Cs$_4$O$_6$ were investigated using a superconducting 
quantum interference device (SQUID, Quantum Design MPMS-XL5). Samples of approximately 100~mg, 
fused in Suprasil tubes under helium atmosphere, were used for the analysis. In the cases
of temperature dependent and time dependent magnetometry we performed  
under zero-field-cooled (ZFC) and field-cooled (FC) measurements. For the ZFC conditions, the
samples were first cooled to a temperature of 1.8~K without applying a magnetic field. After applying 
an induction field $\mu_0H$, the magnetization was recorded as a function of temperature or time.
For the temperature dependent measurements, the magnetization was recorded directly 
afterward in the same field upon lowering the temperature down to 1.8 K again (FC).

\begin{figure}[H]
\includegraphics[width=9cm]{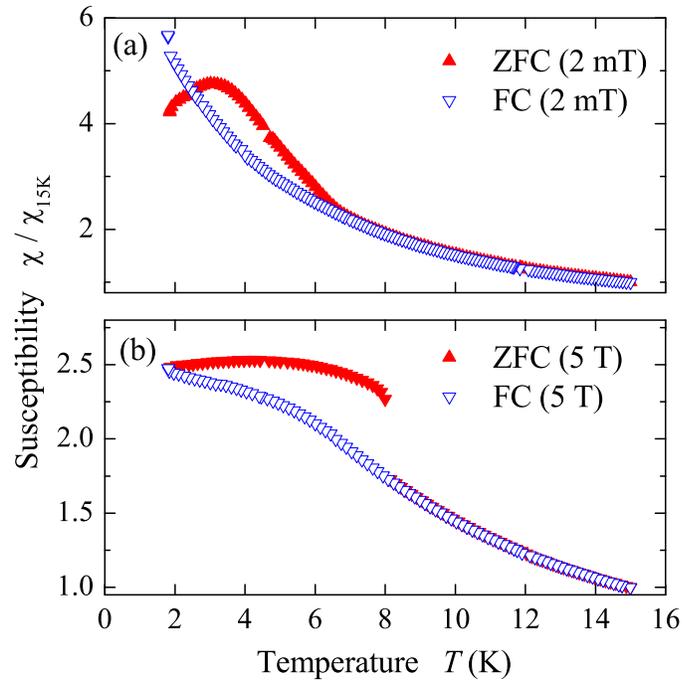}
\caption{(Color online) Shown is the temperature dependent magnetic susceptibility of Cs$_4$O$_6$.
The low-temperature behavior is shown in (a) and (b) for induction fields of 2~mT and 5~T, respectively. 
(All values are normalized by the value of the susceptibility at 2~K.)}
\label{fig_magtempCs}
\end{figure}

Temperature dependent magnetization experiments were discussed in an earlier work
\cite{WFF07} and indicate that Rb$_4$O$_6$ behaves like a frustrated system. We have analyzed
the related Cs$_4$O$_6$ using an identical experimental setup. Figures~\ref{fig_magtempCs}(a) 
and (b) display the temperature dependent magnetization of Cs$_4$O$_6$ at magnetic induction 
fields $\mu_0H$ of 2~mT and 5~T, respectively. The measurements were carried 
out under ZFC conditions and FC conditions. In the 2~mT ZFC measurement, a magnetic transition 
is found to occur at 3.2~$\pm0.2$~K. In the 5~T ZFC curve, this transition shows a distinct 
broadening and is shifted to a higher temperature. Thermal irreversibilities between ZFC and 
FC measurements are observed in both magnetic fields, a behavior known from frustrated systems
and spin glasses. Cs$_4$O$_6$ exhibits similar magnetic properties as Rb$_4$O$_6$.

\begin{figure}[H]
\includegraphics[width=9cm]{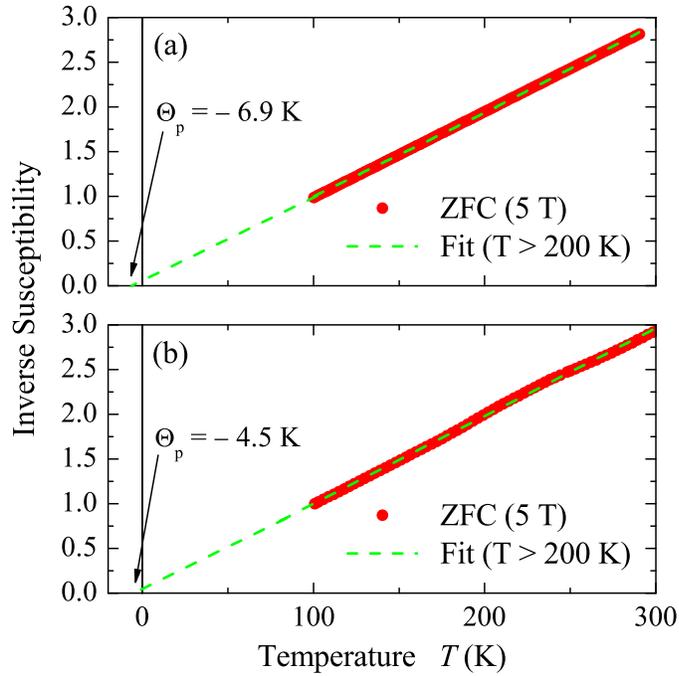}
\caption{(Color online) Curie-Weiss-Fits of Rb$_4$O$_6$~\cite{WFF07} and Cs$_4$O$_6$.
The data of Rb$_4$O$_6$ are shown in (a) and the data of Cs$_4$O$_6$ are shown in (b).
The dashed lines represent a Curie-Weiss-Fit in the temperature
region above 200~K. (All values are normalized by the values of the inverse susceptibilities at 100~K.)}
\label{fig_CWF}
\end{figure}

Figure~\ref{fig_CWF} shows Curie-Weiss-Fits of Rb$_4$O$_6$ and Cs$_4$O$_6$ in  
temperature ranges of 100-300~K at  magnetic fields of 5~T. The fits were performed in
the region above 200~K and yield negative paramagnetic transition temperatures for both
compounds ($\Theta_D$=-6.9~K for Rb$_4$O$_6$~\cite{WFF07} and $\Theta_D$=-4.5~K for Cs$_4$O$_6$).
This indicates dominance of antiferromagnetic interactions. From the high temperature data, 
an effective magnetic moment of $m=1.83\mu_B$ per hyperoxide anion can be deduced applying 
the Curie-Weiss law based on molecular field theory (MFT) for Rb$_4$O$_6$, whereas the moment
was calculated to amount to $m=2.01\mu_B$ for Cs$_4$O$_6$. These values 
are in modest agreement with 1.73~$\mu_B$ as expected from MFT using the spin-only approximation. 
The inverse susceptibility of Cs$_4$O$_6$ exhibits a broad peak at approximately 210~K.
The origin of this peak is not clear. Several alkali oxides are, however, known to 
show multiple phase transitions,\cite{HJS89,RZK78} which are in some cases even very small changes
of the lattice parameters. The peak may correspond to such a phase transition.

\begin{figure}[H]
\includegraphics[width=9cm]{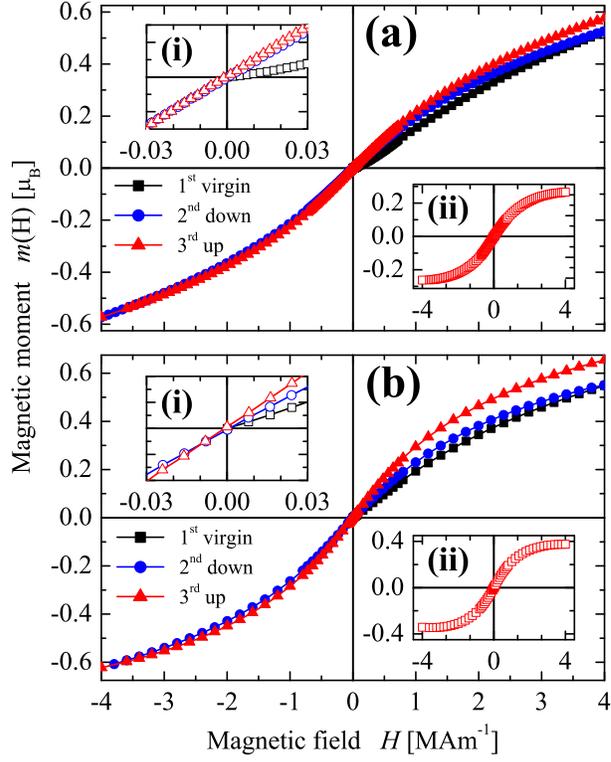}
\caption{(Color online) Field dependent magnetization of
Rb$_4$O$_6$ and Cs$_4$O$_6$ at T = 2~K. Shown are three magnetization
cycles as a function of applied H: virgin ($0 \rightarrow
+H_{max}$), down ($+H_{max} \rightarrow -H_{max}$), and up
($-H_{max} \rightarrow +H_{max}$). The insets (i) show in detail the
magnetization close to the origin. The insets (ii) show the remaining
Langevin functions after subtracting the linear paramagnetic
correction.} 
\label{fig_hys}
\end{figure}

We also performed field-dependent magnetization measurements. Figure~\ref{fig_hys}(a) shows 
the field dependent magnetization of Rb$_4$O$_6$ at 2~K, well below the magnetic transition 
temperature of 3.4~K. It is clear that the magnetization does not show hysteresis 
[see inset (i)]. The shape of the magnetization curve is best modelled by a paramagnetic 
loop correction to a Langevin function $L(H)$ given by $M(H) = \chi_{lin} H + M_0 L(H)$, 
where $\chi_{lin}$ is the field independent paramagnetic susceptibility and $M_0$ the 
saturation moment. In inset (ii), the linear contribution was subtracted from the total 
magnetization. The remaining Langevin function saturates initially with a magnetic moment 
of 0.25~$\mu_B$ per Rb$_4$O$_6$ formula unit and increases in successive cycles. At given 
magnetic fields, the up and down curves exhibit differences in the measured magnetization, 
indicating that the magnetization changes with time in a manner consistent with the known 
relaxation behavior of frustrated systems.\cite{ScW00} Figure~\ref{fig_hys}(b) shows the 
corresponding field dependent magnetization for Cs$_4$O$_6$, qualitatively similar to Rb$_4$O$_6$. 
Applying the model of the paramagnetic loop correction to a Langevin function as 
described above a saturation magnetic moment of 0.37~$\mu_B$ per Cs$_4$O$_6$ formula 
unit is obtained as seen in Figure~\ref{fig_hys}(b) inset (ii). While the total 
magnetic moments of Rb$_4$O$_6$ and Cs$_4$O$_6$ are quite similar at given magnetic 
fields, the paramagnetic contributions exhibit a large difference. This indicates 
that the time dependence and the dynamics of the compounds are differing. 

\begin{figure}
\includegraphics[width=9cm]{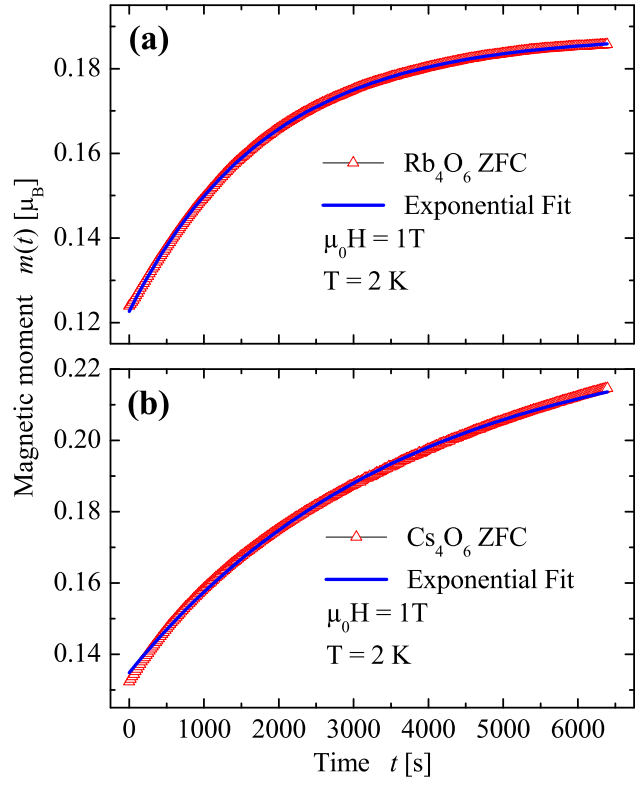}
\caption{(Color online) Time dependent magnetizations of Rb$_4$O$_6$ and Cs$_4$O$_6$
in induction fields of $\mu_0 H = 1$~T are shown. The solid lines are the 
result of exponential fits.} 
\label{fig_mt}
\end{figure}

Time dependent measurements reveal more about these dynamics. Figures~\ref{fig_mt}(a) 
and (b) show the time dependent variations of the magnetization up to 6400~s for both compounds.
The measurements were performed under ZFC conditions at 2~K in induction fields of 1~T. 
The magnetic moment of Rb$_4$O$_6$ varies exponentially with a relaxation time of 
$\tau =(1852 \pm 30)$~s. This value is comparable to those of other frustrated systems. 
Similar curves were obtained for Rb$_4$O$_6$ using lower as well as higher induction 
fields of 30~mT and 5~T. The relaxation times were determined to be $\tau(30$~mT)$=(1170 \pm 30)$~s 
and $\tau(5$~T)$=(4340 \pm 20)$~s. The pronounced relaxation is another clear indication
of the magnetic frustration in Rb$_4$O$_6$. 

As in the case of Rb$_4$O$_6$, the magnetic moment of Cs$_4$O$_6$ follows 
exponential behavior. A relaxation time of $\tau =(3701 \pm 30)$~s 
is deduced from the exponential fit confirming that Cs$_4$O$_6$ is also a magnetically 
frustrated $2p$-system although with different dynamics. The fitting curve does not
follow the exponential behavior as exactly as for Rb$_4$O$_6$. This is most probably due to
the fact that a complete saturation of the magnetization is not reached within the time 
span of 6400~s. A reason for these differences between the compounds cannot 
be given within the scope of these experiments, but electronic structure calculations 
for Cs$_4$O$_6$ may shed more light on the dynamics of this system.

\section{Summary and Conclusions}

It has been shown that the alkali sesquioxides Rb$_4$O$_6$ and Cs$_4$O$_6$ 
exhibit frustrated magnetic ordering based on anionogenic $2p$-electrons of the
hyperoxide anions. Mixed valency was verified in both compounds using Raman spectroscopy.
The strong time dependence of the magnetization and the pronounced differences between 
the ZFC and FC measurements support the previously assumed frustrated state of 
Rb$_4$O$_6$~\cite{WFJ09}. Cs$_4$O$_6$ was found to show a very similar behavior.
The experiments show that strong electronic correlations can also occur in presumably 
simple $2p$-compounds such as alkali oxides. The complex distribution of magnetic moments
in the lattices leads to a symmetry reduction and causes a frustrated magnetic beahvior
in both compounds. These results are of major importance since they confirm that open
shell $p$-electrons can behave like $d$- or $f$-electrons.

\bigskip
\begin{acknowledgments}
This work was funded by the DFG in the Collaborative Research Center
{\it Condensed Matter Systems with Variable Many-Body Interactions}
(TRR 49). The authors are grateful for the fruitful discussions with W.
Pickett, M. Jourdan, G. Jakob and H. von L{\"o}hneysen. T.~P.$^1$,  I.~T.$^2$ and S.~M.$^3$
thank their home institutes for support ($^1$ Institute of Physical Chemistry PAS, Kasprzaka 44/52, 01-224 Warsaw, Poland, $^2$ A.~V. Shubnikov Institute of Crystallography, RAS, 117333, Leninskii pr.59, Moscow, Russia, and $^3$ National Technical University "KhPI", Frunze Str. 21, 61002 Kharkov, Ukraine)
\end{acknowledgments}



\newpage


\newpage

\newpage

\newpage

\newpage

\newpage

\newpage

\newpage


\begin{table}[H]
\caption{Atomic parameters for Rb$_4$O$_6$ and Cs$_4$O$_6$.\\
         a) In the cubic space group $I\:\overline{4}3d$ (no. 220)
         all alkali or oxygen atoms are equivalent. $q = d/(2a)$ is the relative position parameter
         of the oxygen atoms that depends on the bond length $d$ of the dioxygen anions
         and the lattice parameter $a$.\\
         b) In space group $I\:2_12_12_1$ (no. 24)
         each pair O$^{ij}$ ($i=1,2,3$, $j=1,2$) forms one set of O$_2$ anions.
         The centers of these anions are located at (7/8, 0, 1/4),
         (1/4, 7/8, 0), and (0, 1/4, 7/8). $q_1$ and $q_2$ (for values see Table~\ref{tab_qpara})
         are the relative position parameters of the peroxide and hyperoxide anions, respectively.
         All permutations of $q_1$ and $q_2$ lead to the same structure.}
\smallskip
\centering
\begin{ruledtabular}
\begin{tabular}{ l c c c c}
Atom & Site & x & y & z \\
\hline
$I\:\overline{4}3d$  &&& \\
\hline
AM & 16$c$ & t       & t & t \\
O  & 24$d$ & $3/8-q$ & 0 & 3/4 \\
\hline
Rb & 16$c$ & 1.054696(50) & 1.054696(50) & 1.054696(50) \\
O  & 24$d$ & 1.20206(36)  & 0 & 3/4 \\
\hline
\hline
Cs & 16$c$ & 0.946544(45) & 0.946544(45) & 0.946544(45) \\
O  & 24$d$ & 0.55065(49)  & 0 & 3/4 \\
$I\:2_12_12_1$ &&& \\
\hline
        AM$^1$      & 8$d$    & $1/4-t$    & $1/4-t$     & $1/4-t$ \\
        AM$^2$      & 8$d$    & $1/2-t$    & $1/2-t$     & $1/2-t$ \\
        O$^{11}$    & 4$a$    & $7/8-q_2$  & 0           & 1/4     \\
        O$^{12}$    & 4$a$    & $7/8+q_2$  & 0           & 1/4     \\
        O$^{21}$    & 4$b$    & 1/4        & $7/8-q_2$   & 0       \\
        O$^{22}$    & 4$b$    & 1/4        & $7/8+q_2$   & 0       \\
        O$^{31}$    & 4$c$    & 0          & 1/4         & $7/8-q_1$ \\
        O$^{32}$    & 4$c$    & 0          & 1/4         & $7/8+q_1$ \\
\end{tabular}
\end{ruledtabular}
\label{tab_xrd}
\end{table}

\begin{table}[H]
    \caption{Structural parameters for the alkali sesquioxides AM$_4$O$_6$.\\
             Cs data from~\cite{HeK393};
             Rb data from~\cite{JaK91,JHK99} for powder XRD (pd), single crystal (sc) XRD
             and powder neutron diffraction (nd).
             $q=d/(2a)$ is the mean parameter from XRD measurements,
             $q_2$ and $q_1$ are the parameters calculated
             for O$_2^-$ and O$_2^{--}$ ions, respectively (see text).}
        \begin{ruledtabular}
        \begin{tabular}{ l | ccccccc}
        AM         & $a$~[\AA] & $t$     & $d$~[\AA] & $q$     & $q_2$  & $q_1$  & method \\
        \hline
        Rb         & 9.3242   & 0.0545  & 1.3426   & 0.072   &        &        & pd \\
        Rb (295~K) & 9.3327   & 0.05411 & 1.3234   & 0.0709  &        &        & sc \\
        Rb (213~K) & 9.2884   & 0.05377 & 1.3115   & 0.0706  & 0.0721 & 0.0829 & sc \\
        Rb (5~K)   & 9.2274   & 0.05395 & 1.363    & 0.0738  & 0.0726 & 0.0834 & nd \\
        Cs         & 9.86     & 0.054   &          &         & 0.068  & 0.078  & pd \\
        \end{tabular}
        \end{ruledtabular}
    \label{tab_qpara}
\end{table}

\end{document}